\documentclass[twocolumn,showpacs,preprintnumbers,amsmath,amssymb]{revtex4}
\usepackage{graphicx}
\usepackage{dcolumn}
\usepackage{bm}

\begin{document}

\preprint{APS/123-QED}

\title{Frequency filtered storage of parametric fluorescence\\with electromagnetically induced transparency}
\author{K. Akiba$^1$}
\author{K. Kashiwagi$^1$}
\author{T. Yonehara$^1$}
\author{M. Kozuma$^{1,2}$} 
\affiliation{%
$^1$Department of Physics, Tokyo Institute of Technology, 
2-12-1 O-okayama, Meguro-ku, Tokyo 152-8550, Japan \\
$^2$PRESTO, CREST, Japan Science and Technology Agency, 1-9-9 Yaesu, Chuo-ku, Tokyo 103-0028, Japan
}
\date{\today}             

\begin{abstract}
The broadband parametric fluorescence pulse (probe light) with center frequency resonant on $^{87}$Rb $D_1$ line 
was injected into a cold atomic ensemble with coherent light (control light). 
Due to the low gain in the parametric down conversion process, the probe light was in a highly bunched photon-pair state.
By switching off the control light, the probe light within the electromagnetically induced transparency window was mapped on the atoms. 
When the control light was switched on, 
the probe light was retrieved and frequency filtered storage was confirmed from the superbunching effect
and an increase of the coherence time of the retrieved light. 
\end{abstract}

\maketitle

Dark state polariton theory \cite{DSP,QM} of electromagnetically induced transparency (EIT) \cite{EIT1} presents the possibility of coherently transferring 
a quantum state of light to an atomic collective excitation by manipulating control light,  where the control light turns an opaque medium for a weak probe light into a transparent one. 
Such a quantum state transfer has been successfully demonstrated for nonclassical light 
generated from an atomic ensemble \cite{Kuzmich, Lukin, remote}. The advantage of generating nonclassical light from the atoms is that the bandwidth of the light is matched by the narrow atomic resonant width. 
Compared to the process of generating nonclassical lights with atoms, 
the parametric down conversion (PDC) process has been widely used in fundamental 
experiments in quantum optics \cite{Squeeze,Mandel,QN2} and 
for demonstrating various 
quantum information protocols \cite{QT,QCR}. 
The process also enables the generation of rich nonclassical photonic states such as 
the antibunched state \cite{Koashi,Ou}, 
hyperentangled state \cite{Hyperentangle}
GHZ state \cite{GHZ}, 
$W$ state \cite{W},
and cluster state \cite{cluster}. 
Storing such nonclassical light in atomic ensembles 
generates corresponding nonclassical atomic states. 
Furthermore, the state of the nonclassical light can be manipulated by applying an additional field to the atoms storing the photonic state or by changing the properties of the control light \cite{manipulate1,manipulate2,SoL2}.

In this paper, we report frequency filtered storage of parametric fluorescence (highly bunched photon-pair state) with EIT.
The frequency bandwidth of parametric fluorescence is typically of the order of $10^{13}$ Hz, 
which is much broader than an atomic resonance interaction width ($\sim10^{6}$ Hz). 
The discrepancy between the frequency bandwidth and the resonance width presents a problem 
for photon counting in experiments, 
since photons interacting with atoms cannot be selectively counted. 
We avoided this problem in our experiments by utilizing the properties of the optical pulse propagation inside the EIT medium \cite{DSP,QM,timefrequency}.
When the pulse spectrum was broader than the transparency window,
the pulse was split into 
an adiabatic part having a spectrum within the transparency window 
and 
an oscillating nonadiabatic part.
The control light was turned off when the pulse was propagating in the medium, thus storing the adiabatic part. 
After the nonadiabatic part passed through the medium, the control light was turned on, allowing the retrieved adiabatic part to be separated from the nonadiabatic part in time. We call this technique as time-frequency filtering with storage of light. The disadvantage of this technique is that the frequency bandwidth of the retrieved light is limited to that of the EIT transparency window and the count rate after the storage and retrieval is decreased by a factor equal to the ratio of the EIT width to the incident light bandwidth.

\begin{figure}
 \includegraphics{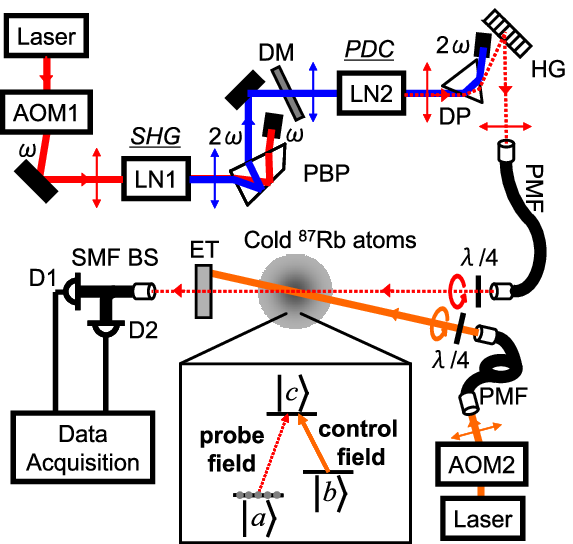}
  \caption{\label{fig:expsetup}(Color online) 
  Schematic diagram of experimental setup. 
  AOM: acousto-optic modulator; LN: quasi-phase-matched MgO:LiNbO$_3$ waveguide; 
  PBP: Pellin Broca prism; DM: dichroic mirror; DP: dispersing prism; HG: holographic grating; 
  PMF: polarization maintained single mode fiber; $\lambda$/4: quarter wave plate; ET: etalon; 
  BS: beam splitter; SMF: single mode fiber; D1, D2: avalanche photodetectors.  
}

\end{figure}
Fig. \ref{fig:expsetup} shows a schematic diagram of the experimental setup.
The states \{$|{a}\rangle$, $|{b}\rangle$\} and $|{c}\rangle
$ of a $\Lambda$ type three level system 
correspond to $5^2S_{1/2},\ F=\{1, 2\}$ and $5^2P_{1/2},\ F'=2$ of the $^{87}$Rb $D_1$ line, respectively. 
We generated parametric fluorescence by using two independent quasi-phase-matched MgO:LiNbO$_3$ waveguides (LN1,2 in Fig.~\ref{fig:expsetup}) \cite{SVEIT,FFPF}.
Optical pulses (pulse width: 190 ns; peak intensity: $\sim$110 mW) were generated by an acousto-optic modulator (AOM1) from the linearly polarized continuous output of a Ti:sapphire laser.
The pulses were coupled to LN1 with 50\% efficiency and generated second harmonic pulses (pulse width: 130 ns; peak intensity: $\sim$16 mW). 
The second harmonic light was injected into LN2 (coupling efficiency 40\%) and
the parametric fluorescence was generated. 
The frequency of the degenerate component in the parametric fluorescence 
was resonant with the $|{a}\rangle\to|{c}\rangle$ transition. 
The frequency bandwidth of the fluorescence was about 10 THz, which was reduced to 23 GHz using a holographic grating (HG) and a polarization maintained single mode fiber (PMF). 
The output of the PMF was used as broadband probe light.
In order to obtain a high extinction ratio between the probe and control lights, the two beams were overlapped at the cold atoms with a crossing angle of 2.5$^\circ$ and an etalon (ET) was inserted into the probe optical axis. The frequency width of the parametric fluorescence was thus reduced to about 600 MHz before detection. 
The overall transmission efficiency for the degenerate component of the probe light without the atoms was about 3\%.

We first evaluated the intensity correlation function of the obtained broadband probe light.
The light was detected by photon counting using silicon avalanche photodiodes 
(D1, D2; PerkinElmer model SPCM-AQR; detection efficiency: 62\%; time resolution: 350 ps).
The outputs of D1 and D2 were fed to the start and stop inputs, respectively, of a multi-channel scaler (MCS).
\begin{figure}
 \includegraphics{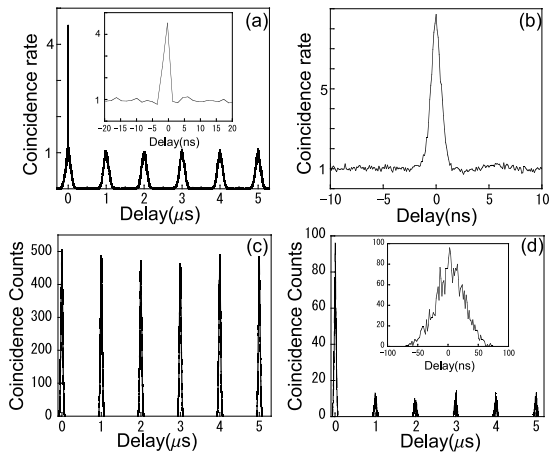}
  \caption{\label{fig:g2} 
  Experimental results of intensity correlation with 1.6-ns resolution of the MCS:
  (a) broadband parametric fluorescence,
  (c) retrieved component of coherent light, and 
  (d) retrieved component of parametric fluorescence.
  The insets of (a) and (d) show magnifications of the coincidences at 0 time delay.
  (b) Sharp coincidence peak in (a) measured with 0.1 ns resolution of the MCS. 
  }
\end{figure}
Fig. ~\ref{fig:g2}(a) shows the normalized coincidence rate as a function of time delay with 1.6 ns resolution. 
A sharp coincidence peak superposed on a moderate profile of width 0.2 $\mu$s appears at 0 $\mu$s, shown magnified in the inset. 
The moderate profile of the coincidence peak is due to
the shape of the parametric fluorescence pulse. 
The obtained normalized auto intensity correlation function was $g^{(2)}(0)=4.4\pm0.2>3$
\cite{FFPF} 
, which is called ``superbunching''.  
We remeasured the sharp coincidence peak at 0.1 ns resolution (Fig.~\ref{fig:g2}(b)) to closely evaluate the coherence time of the probe light.
The temporal width of the coincidence (full width at half maximum: FWHM) was 1.2 ns with total time resolution 0.8 ns, which 
means the coherence time of the probe light is about or less than 0.2 ns. The auto intensity correlation function is theoretically given by $g^{(2)}(0)=37$ with a squeeze parameter of 0.17 obtained from the classical parametric gain \cite{Loudon}, which is much larger than the experimental value.
This disagreement can be explained by the longer time resolution (2.3 ns) than the coherence time ($\le 0.2$ ns) and fluctuation of the second harmonic light intensity \cite{FFPF}. 
We note that the superbunching effect is not direct proof of the nonclassicality of the light. However, when $g^{(2)}(0)$ is much greater than 3, the state of the parametric fluorescence can be approximated as a highly bunched photon-pair state i.e. $ |{\phi}\rangle \approx |{0}\rangle + \epsilon |{2}\rangle \ (|\epsilon |^2 \ll 1)$.
In our experiment, we prepared the parametric fluorescence such that the fluorescence was in the highly bunched state.

Before performing an experiment with broadband parametric fluorescence, we performed an EIT experiment 
using narrowband coherent light (spectrum line width $<$ 100 kHz) as probe light.
Magneto-optically trapped $^{87}$Rb atoms were used as an optically thick medium, 
placed in a vacuum cell magnetically shielded by a single permalloy.
One cycle of the experiment comprised
an atomic medium preparation period (6.3 ms) and a measurement period (3.7 ms). 
After laser cooling, 
the magnetic fields and the cooling and repumping lights were turned off, and depumping lights  illuminated the atomic ensemble for 45 $\mu$s, so that all atoms were prepared in state $|{a}\rangle$, where the optical depth of the $|{a}\rangle\to|{c}\rangle$ transition was $\sim$6.
In the measurement period, a weak (about 1 pW) probe light (beam diameter 330 $\mu$m) and a 400 $\mu$W control light (beam diameter 770 $\mu$m) with the same circular polarizations were injected into the unpolarized atomic ensemble.
The control light was generated from a second Ti:sapphire laser and 
its intensity was varied using an acousto-optic module (AOM2). 

\begin{figure}
  \includegraphics[width=0.28\textwidth]{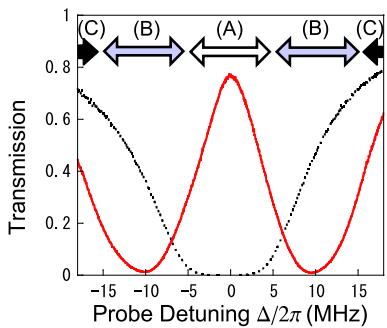}
  \caption{\label{fig:EIT} (Color online)
  Measured transmission spectra of a weak coherent probe 
  as a function of detuning
  with (solid line) and without (dotted line) the control light.
   }
\end{figure}
Fig. ~\ref{fig:EIT} shows probe intensity transmissions as a function of the detuning $\Delta/2\pi$,  
where signals were obtained by sweeping the probe frequency during the measurement period.
In the absence of the control light (dotted line), 
the probe light was absorbed by the atomic medium. 
With the addition of the control light (solid line), 
the medium was rendered transparent around $\Delta/2\pi=0$ 
(peak transmission: 77\%; FWHM of the transparency window: 8.3 MHz). 
We classified this EIT spectrum into three regions: (A) transparent, (B) absorption, and (C) off resonant regions. 
The broadband parametric fluorescence was spread over all regions 

We observed the propagation of the coherent probe pulse in the three regions 
by injecting probe pulses into the cold atoms 3390 times over the measurement period. 
\begin{figure}
 \includegraphics{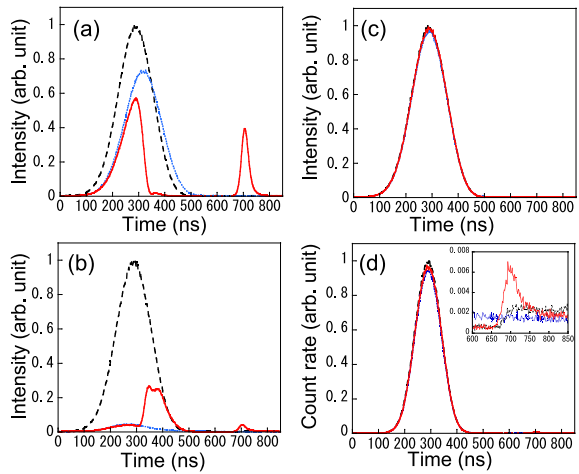}
  \caption{\label{fig:Storage}(Color online) 
     Obtained probe pulses 
     under reference (broken line), slow propagation (dotted line), 
     and storage conditions (solid line). See text for details..
     The probe light was coherent light with frequency detuning of 
     (a) 0 MHz, (b) 10 MHz, and (c) 100 MHz.
     (d) The probe light was broadband parametric fluorescence .
     The inset of (d) shows magnification of the count rate from 600 ns to 850 ns.  
     }
\end{figure}
Figs. ~\ref{fig:Storage}(a)--(c) 
show typical probe signals in the (A)--(C) regions.
The broken curves correspond to transmitted probe intensity signals without cold atoms, where the intensity of 
the control light was dynamically changed (reference condition). The dotted curves correspond to probe signals with cold atoms and constant control light (slow propagation condition). The solid curves show signals with cold atoms and control light with dynamically varied intensity (storage condition). The control light (intensity: 400 $\mu$W) was turned off from 300 ns to 350 ns and turned on from 650 ns to 700 ns. 
For probe detuning of $\Delta/2\pi=$ 0 ((A) tranparent region), a pulse delay of 25 ns was observed under the slow propagation condition (dotted curve in (a)), which corresponds to more than three orders of magnitude reduction in group velocity. Under the storage condition, the probe pulse was partially stored and retrieved (solid curve). 
When the probe frequency was detuned by $\Delta/2\pi=$ 10 MHz (absorption (B) region),
$-$10 ns of ``superluminal'' propagation was observed (dotted curve in (b)), due to anomalous dispersion corresponding to the absorption. Under the storage condition, 
the probe intensity increased when the control light was tuned off and a slight retrieval was observed when the control light was turned on (solid curve in (b)). The increase of the probe intensity can be explained as being due to a change in the probe transmission from the solid to the dotted curve at $\Delta/2\pi=$ 10 MHz in Fig.~\ref{fig:EIT}. 
The slight retrieval will be due to the frequency component of the probe corresponding to the transparent (A) region, since the frequency width of a light pulse is given by the inverse of its temporal width. 
When the probe was detuned by $\Delta/2\pi=$ 100 MHz, the probe signals were almost identical under each of the conditions (Fig.~\ref{fig:Storage}(c)). 
Fig. ~\ref{fig:g2}(c) shows the intensity correlation 
of the retrieved component in Fig.~\ref{fig:Storage}(a), 
where the outputs of D1 and D2 were gated from 675 ns to 745 ns.
The obtained intensity correlation function $g^{(2)}(0) = 1.05\pm0.01$ agrees with that of coherent light ($g^{(2)}(0) = 1$) \cite{abscase}.

Finally, we observed the propagation of broadband parametric fluorescence under EIT, where the center frequency of the fluorescence was set to the resonant frequency corresponding to $|{a}\rangle\to|{c}\rangle$ transition.
Figs. ~\ref{fig:Storage}(d) shows photon count rates as a function of time under reference, slow propagation, and storage conditions.
The inset shows magnifications of the count rates from 600 ns to 850 ns.  
The experimental conditions were the same as for the case of the probe light in the coherent state. 
Since the frequency width of the probe light was broad, 
corresponding phenomena from each of the regions occurred simultaneously.  
There was almost no difference between the reference and the slow propagation conditions, 
since the dominant frequency components of the probe were far from resonance.  
However, storage condition was slightly different from the other conditions. 
There was a small retrieved component from 650 ns to 750 ns. 
The broken and dotted lines in the inset of (d) show residual control light under reference and slow propagation conditions, respectively. Almost all components of the residual control light were caused by the diffuse reflection from the vacuum cell. 
The solid line in the inset of (d) (storage condition) shows the clear retrieved component. (We checked that the retrieved component disappeared in the absence of the probe light.)
The gross count of the retrieved component (the offset due to the residual control light was subtracted) was 0.12\% of that of the incident broadband probe light (in (d) broken line), which agrees with the product of the frequency width reduction (8.3 MHz/600 MHz) and the retrieval efficiency (8\%) of the coherent light in (a).   

Fig. \ref{fig:g2}(d) shows the intensity correlation of the retrieved component. 
The obtained normalized auto intensity correlation function was $g^{(2)}(0) = 9.0\pm0.2 >3$, 
i.e.\ superbunching was obtained, where the coincidence count rate was 0.4 s$^{-1}$ and the experimental time was 2 hours.
The coincidence at 0 time delay in Fig.~\ref{fig:g2}(a) has a sharp peak superposed on a moderate profile, while that in (d) does not have such a double structure, where the signals were obtained under the same time resolution. 
The observed coherence time of the retrieved light was $\sim$35 ns, which is much longer than that of the incident probe pulse. 
The increase of the coherence time shows that frequency filtered storage and retrieval occurred, that is, 
we have demonstrated the time-frequency filtering with storage of light technique. 
The experimentally obtained $g^{(2)}(0)$ should be equal to the theoretical value of 37, 
since the coherence time was larger than the time resolution for detection.
The disagreement can be expalained as follows. 
The ratio of the retrieved light to the residual control light was almost unity in the gated time. Mixing of the parametric fluorescence ($g^{(2)}(0)=37$) with the coherent light($g^{(2)}(0)=1$) by 1 to 1 ratio gives the theoretical value of $g^{(2)}(0)=11$, which well explains the experimentally obtained value ($9.0\pm0.2$). The slight difference can be explained by the fluctuation of the second harmonic light intensity (squeeze parameter: $r$).

\begin{figure}
 \includegraphics{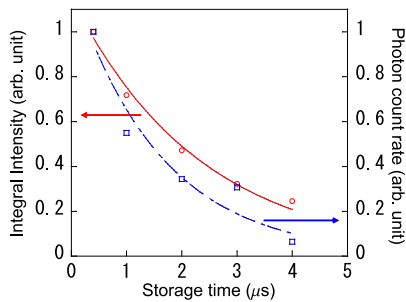}
  \caption{\label{fig:CT}(Color online) 
  Integral intensities of retrieved coherent lights and 
  photon count rates of retrieved parametric fluorescence as a function of storage time.
     }
\end{figure}
Fig. ~\ref{fig:CT} shows 
the integral intensities of the retrieved coherent light and 
photon count rates of the retrieved parametric fluorescence (the offset due to the residual control light was subtracted) as a function of storage time. 
The obtained $1/e$ decay times were 
2.3 $\mu$s and 1.6 $\mu$s for coherent light and parametric fluorescence, respectively. 
Signal decay was the result of the destruction of atomic coherence, which was in our experiment caused by residual magnetic fields due to eddy currents.  
The small difference in the atomic coherence times  
may be caused by a disturbance of the collective atomic state due to absorptions of the broadband parametric fluorescence in the absorption (B) region . 

In conclusion, we have demonstrated frequency filtered storage of broadband parametric fluorescence (highly bunched photon-pair state).
The frequency component of the fluorescence within an electromagnetically induced transparency window was stored in an atomic ensemble and then retrieved after other components had passed through the atoms, confirmed by the observation of superbunching and from evaluating the coherence time for the retrieved light. 
This experiment also demonstrated the time-frequency filtering with storage of light technique. 
While the observation of the superbunching effect is not direct proof of the nonclassicality of the light, 
the results obtained here are the first steps toward 
the generation of various nonclassical states of atomic ensembles 
with the widely used parametric down conversion process.  

We gratefully acknowledge R. Inoue, D. Akamatsu and Y. Yokoi.
One of the authors (K. A.) was partially supported by the JSPS.
This work was supported by a Grant-in-Aid for Scientific Research (B) and 
the 21st Century COE Program at Tokyo Tech ``Nanometer-Scale Quantum Physics'' by MEXT.

\bibliography{paper}

\begin{thebibliography}{25}
\expandafter\ifx\csname natexlab\endcsname\relax\def\natexlab#1{#1}\fi
\expandafter\ifx\csname bibnamefont\endcsname\relax
  \def\bibnamefont#1{#1}\fi
\expandafter\ifx\csname bibfnamefont\endcsname\relax
  \def\bibfnamefont#1{#1}\fi
\expandafter\ifx\csname citenamefont\endcsname\relax
  \def\citenamefont#1{#1}\fi
\expandafter\ifx\csname url\endcsname\relax
  \def\url#1{\texttt{#1}}\fi
\expandafter\ifx\csname urlprefix\endcsname\relax\def\urlprefix{URL }\fi
\providecommand{\bibinfo}[2]{#2}
\providecommand{\eprint}[2][]{\url{#2}}

\bibitem[{\citenamefont{Fleischhauer and Lukin}(2000)}]{DSP}
\bibinfo{author}{\bibfnamefont{M.}~\bibnamefont{Fleischhauer}}
  \bibnamefont{and} \bibinfo{author}{\bibfnamefont{M.~D.} \bibnamefont{Lukin}},
  \bibinfo{journal}{Phys.\ Rev.\ Lett.} \textbf{\bibinfo{volume}{84}},
  \bibinfo{pages}{5094} (\bibinfo{year}{2000}).

\bibitem[{\citenamefont{Fleischhauer and Lukin}(2002)}]{QM}
\bibinfo{author}{\bibfnamefont{M.}~\bibnamefont{Fleischhauer}}
  \bibnamefont{and} \bibinfo{author}{\bibfnamefont{M.~D.} \bibnamefont{Lukin}},
  \bibinfo{journal}{Phys.\ Rev.\ A} \textbf{\bibinfo{volume}{65}},
  \bibinfo{pages}{022314} (\bibinfo{year}{2002}).

\bibitem[{\citenamefont{Harris}(1997)}]{EIT1}
\bibinfo{author}{\bibfnamefont{S.~E.} \bibnamefont{Harris}},
  \bibinfo{journal}{Phys.\ Today} \textbf{\bibinfo{volume}{50 {\textmd{(7)}}}},
  \bibinfo{pages}{36} (\bibinfo{year}{1997}).

\bibitem[{\citenamefont{Chaneli\`{e}re\textit{\ et al}}(2005)}]{Kuzmich}
\bibinfo{author}{\bibfnamefont{T.}~\bibnamefont{Chaneli\`{e}re\textit{\ et
  al}}}, \bibinfo{journal}{Nature (London)} \textbf{\bibinfo{volume}{438}},
  \bibinfo{pages}{833} (\bibinfo{year}{2005}).

\bibitem[{\citenamefont{Eisaman\textit{\ et al}}(2005)}]{Lukin}
\bibinfo{author}{\bibfnamefont{M.~D.} \bibnamefont{Eisaman\textit{\ et al}}},
  \bibinfo{journal}{Nature (London)} \textbf{\bibinfo{volume}{438}},
  \bibinfo{pages}{837} (\bibinfo{year}{2005}).

\bibitem[{\citenamefont{Matsukevich\textit{\ et al}}(2006)}]{remote}
\bibinfo{author}{\bibfnamefont{D.~N.} \bibnamefont{Matsukevich\textit{\ et
  al}}}, \bibinfo{journal}{Phys.\ Rev.\ Lett.} \textbf{\bibinfo{volume}{96}},
  \bibinfo{pages}{030405} (\bibinfo{year}{2006}).

\bibitem[{\citenamefont{Wu et~al.}(1986)\citenamefont{Wu, Kimble, Hall, and
  Wu}}]{Squeeze}
\bibinfo{author}{\bibfnamefont{L.~A.} \bibnamefont{Wu}},
  \bibinfo{author}{\bibfnamefont{H.~J.} \bibnamefont{Kimble}},
  \bibinfo{author}{\bibfnamefont{J.~L.} \bibnamefont{Hall}}, \bibnamefont{and}
  \bibinfo{author}{\bibfnamefont{H.}~\bibnamefont{Wu}},
  \bibinfo{journal}{Phys.\ Rev.\ Lett.} \textbf{\bibinfo{volume}{57}},
  \bibinfo{pages}{2520} (\bibinfo{year}{1986}).

\bibitem[{\citenamefont{Mandel}(1999)}]{Mandel}
\bibinfo{author}{\bibfnamefont{L.}~\bibnamefont{Mandel}},
  \bibinfo{journal}{Rev.\ Mod.\ Phys.} \textbf{\bibinfo{volume}{71}},
  \bibinfo{pages}{S274} (\bibinfo{year}{1999}).

\bibitem[{\citenamefont{Kwiat\textit{\ et al}}(1995)}]{QN2}
\bibinfo{author}{\bibfnamefont{P.~G.} \bibnamefont{Kwiat\textit{\ et al}}},
  \bibinfo{journal}{Phys.\ Rev.\ Lett.} \textbf{\bibinfo{volume}{75}},
  \bibinfo{pages}{4337} (\bibinfo{year}{1995}).

\bibitem[{\citenamefont{Bouwmeester\textit{\ et al}}(1997)}]{QT}
\bibinfo{author}{\bibfnamefont{D.}~\bibnamefont{Bouwmeester\textit{\ et al}}},
  \bibinfo{journal}{Nature (London)} \textbf{\bibinfo{volume}{390}},
  \bibinfo{pages}{575} (\bibinfo{year}{1997}).

\bibitem[{\citenamefont{Gisin\textit{\ et al}}(2002)}]{QCR}
\bibinfo{author}{\bibfnamefont{N.}~\bibnamefont{Gisin\textit{\ et al}}},
  \bibinfo{journal}{Rev.\ Mod.\ Phys.} \textbf{\bibinfo{volume}{74}},
  \bibinfo{pages}{145} (\bibinfo{year}{2002}).

\bibitem[{\citenamefont{Koashi et~al.}(1990)\citenamefont{Koashi, Matsuoka, and
  Hirano}}]{Koashi}
\bibinfo{author}{\bibfnamefont{M.}~\bibnamefont{Koashi}},
  \bibinfo{author}{\bibfnamefont{M.}~\bibnamefont{Matsuoka}}, \bibnamefont{and}
  \bibinfo{author}{\bibfnamefont{T.}~\bibnamefont{Hirano}},
  \bibinfo{journal}{Phys.\ Rev.\ A} \textbf{\bibinfo{volume}{53}},
  \bibinfo{pages}{3621} (\bibinfo{year}{1990}).

\bibitem[{\citenamefont{Lu and Ou}(2001)}]{Ou}
\bibinfo{author}{\bibfnamefont{Y.~J.} \bibnamefont{Lu}} \bibnamefont{and}
  \bibinfo{author}{\bibfnamefont{Z.~Y.} \bibnamefont{Ou}},
  \bibinfo{journal}{Phys.\ Rev.\ Lett.} \textbf{\bibinfo{volume}{88}},
  \bibinfo{pages}{023601} (\bibinfo{year}{2001}).

\bibitem[{\citenamefont{Barreiro et~al.}(2005)\citenamefont{Barreiro, Langford,
  Peters, and Kwiat}}]{Hyperentangle}
\bibinfo{author}{\bibfnamefont{J.~T.} \bibnamefont{Barreiro}},
  \bibinfo{author}{\bibfnamefont{N.~K.} \bibnamefont{Langford}},
  \bibinfo{author}{\bibfnamefont{N.~A.} \bibnamefont{Peters}},
  \bibnamefont{and} \bibinfo{author}{\bibfnamefont{P.~G.} \bibnamefont{Kwiat}},
  \bibinfo{journal}{Phys. Rev. Lett.} \textbf{\bibinfo{volume}{95}},
  \bibinfo{pages}{260501} (\bibinfo{year}{2005}).

\bibitem[{\citenamefont{Bouwmeester et~al.}(1999)\citenamefont{Bouwmeester,
  Pan, Daniell, Weinfurter, and Zeilinger}}]{GHZ}
\bibinfo{author}{\bibfnamefont{D.}~\bibnamefont{Bouwmeester}},
  \bibinfo{author}{\bibfnamefont{J.~W.} \bibnamefont{Pan}},
  \bibinfo{author}{\bibfnamefont{M.}~\bibnamefont{Daniell}},
  \bibinfo{author}{\bibfnamefont{H.}~\bibnamefont{Weinfurter}},
  \bibnamefont{and}
  \bibinfo{author}{\bibfnamefont{A.}~\bibnamefont{Zeilinger}},
  \bibinfo{journal}{Phys. Rev. Lett.} \textbf{\bibinfo{volume}{82}},
  \bibinfo{pages}{1345} (\bibinfo{year}{1999}).

\bibitem[{\citenamefont{Eibl et~al.}(2004)\citenamefont{Eibl, Kiesel,
  Bourennane, Kurtsiefer, and Weinfurter}}]{W}
\bibinfo{author}{\bibfnamefont{M.}~\bibnamefont{Eibl}},
  \bibinfo{author}{\bibfnamefont{N.}~\bibnamefont{Kiesel}},
  \bibinfo{author}{\bibfnamefont{M.}~\bibnamefont{Bourennane}},
  \bibinfo{author}{\bibfnamefont{C.}~\bibnamefont{Kurtsiefer}},
  \bibnamefont{and}
  \bibinfo{author}{\bibfnamefont{H.}~\bibnamefont{Weinfurter}},
  \bibinfo{journal}{Phys. Rev. Lett.} \textbf{\bibinfo{volume}{92}},
  \bibinfo{pages}{077901} (\bibinfo{year}{2004}).

\bibitem[{\citenamefont{Walther{\textit{\ et al}}}(2005)}]{cluster}
\bibinfo{author}{\bibfnamefont{P.}~\bibnamefont{Walther{\textit{\ et al}}}},
  \bibinfo{journal}{Nature (London)} \textbf{\bibinfo{volume}{434}},
  \bibinfo{pages}{169} (\bibinfo{year}{2005}).

\bibitem[{\citenamefont{Akamatsu and Kozuma}(2003)}]{manipulate1}
\bibinfo{author}{\bibfnamefont{D.}~\bibnamefont{Akamatsu}} \bibnamefont{and}
  \bibinfo{author}{\bibfnamefont{M.}~\bibnamefont{Kozuma}},
  \bibinfo{journal}{Phys. Rev. A} \textbf{\bibinfo{volume}{67}},
  \bibinfo{pages}{023803} (\bibinfo{year}{2003}).

\bibitem[{\citenamefont{Patnaik et~al.}(2004)\citenamefont{Patnaik, Kien, and
  Hakuta}}]{manipulate2}
\bibinfo{author}{\bibfnamefont{A.~K.} \bibnamefont{Patnaik}},
  \bibinfo{author}{\bibfnamefont{F.~L.} \bibnamefont{Kien}}, \bibnamefont{and}
  \bibinfo{author}{\bibfnamefont{K.}~\bibnamefont{Hakuta}},
  \bibinfo{journal}{Phys. Rev. A} \textbf{\bibinfo{volume}{69}},
  \bibinfo{pages}{035803} (\bibinfo{year}{2004}).

\bibitem[{\citenamefont{Liu\textit{\ et al}}(2001)}]{SoL2}
\bibinfo{author}{\bibfnamefont{C.}~\bibnamefont{Liu\textit{\ et al}}},
  \bibinfo{journal}{Nature (London)} \textbf{\bibinfo{volume}{409}},
  \bibinfo{pages}{490} (\bibinfo{year}{2001}).

\bibitem[{\citenamefont{Shakhmuratov and Odeurs}(2005)}]{timefrequency}
\bibinfo{author}{\bibfnamefont{R.~N.} \bibnamefont{Shakhmuratov}}
  \bibnamefont{and} \bibinfo{author}{\bibfnamefont{J.}~\bibnamefont{Odeurs}},
  \bibinfo{journal}{Phys. Rev. A} \textbf{\bibinfo{volume}{71}},
  \bibinfo{pages}{013819} (\bibinfo{year}{2005}).

\bibitem[{\citenamefont{Akamatsu et~al.}(2004)\citenamefont{Akamatsu, Akiba,
  and Kozuma}}]{SVEIT}
\bibinfo{author}{\bibfnamefont{D.}~\bibnamefont{Akamatsu}},
  \bibinfo{author}{\bibfnamefont{K.}~\bibnamefont{Akiba}}, \bibnamefont{and}
  \bibinfo{author}{\bibfnamefont{M.}~\bibnamefont{Kozuma}},
  \bibinfo{journal}{Phys. Rev. Lett.} \textbf{\bibinfo{volume}{92}},
  \bibinfo{pages}{203602} (\bibinfo{year}{2004}).

\bibitem[{\citenamefont{Akiba et~al.}(2006)\citenamefont{Akiba, Akamatsu, and
  Kozuma}}]{FFPF}
\bibinfo{author}{\bibfnamefont{K.}~\bibnamefont{Akiba}},
  \bibinfo{author}{\bibfnamefont{D.}~\bibnamefont{Akamatsu}}, \bibnamefont{and}
  \bibinfo{author}{\bibfnamefont{M.}~\bibnamefont{Kozuma}},
  \bibinfo{journal}{Opt. Commun.} \textbf{\bibinfo{volume}{259}},
  \bibinfo{pages}{789} (\bibinfo{year}{2006}).

\bibitem[{\citenamefont{Loudon}(2000)}]{Loudon}
\bibinfo{author}{\bibfnamefont{R.}~\bibnamefont{Loudon}},
  \emph{\bibinfo{title}{The Quantum Theory of Light}}
  (\bibinfo{publisher}{Oxford University Press}, \bibinfo{address}{New York},
  \bibinfo{year}{2000}), \bibinfo{edition}{3rd} ed.

\bibitem[{abs()}]{abscase}
\bibinfo{note}{The intensity correlation function of the retreived component in
  Fig. \ref{fig:Storage}(b) was measured to be $g^{(2)}(0) = 1.06 \pm 0.01$.}

\end{thebibliography}
\end{document}